\documentclass[%
preprint,
superscriptaddress,
onecolumn,
amsmath,amssymb,
aps,
showkeys,
notitlepage,
]{revtex4-1}

\usepackage{graphicx}% Include figure files
\usepackage{dcolumn}% Align table columns on decimal point
\usepackage{bm}% bold math
\usepackage[final]{changes}
\definechangesauthor[name={Victoria Norman},color=red]{VN}
\definechangesauthor[name={Sridhar Majety}, color=olive]{SM}
\definechangesauthor[name={Zhipan Wang},color=orange]{ZW}
\definechangesauthor[name={William Casey},color=yellow]{WC}
\definechangesauthor[name={Nicholas Curro}, color=purple]{NC}
\definechangesauthor[name={Marina Radulaski}, color=blue]{MR}

\begin{document}
	% \preprint{APS/123-QED}
	
	\title{Novel color center platforms enabling fundamental scientific discovery}

	\author{Victoria A. Norman}
	\affiliation{Department of Electrical and Computer Engineering, University of California at Davis, One Shields Ave., Davis, CA 95616, USA}
	\affiliation{Department of Physics, University of California at Davis, One Shields Ave., Davis, CA 95616, USA}
	\author{Sridhar Majety}
	\affiliation{Department of Electrical and Computer Engineering, University of California at Davis, One Shields Ave., Davis, CA 95616, USA}
	\author{Zhipan Wang}
	\affiliation{Department of Physics, University of California at Davis, One Shields Ave., Davis, CA 95616, USA}
	\author{William H. Casey}
	\affiliation{Department of Chemistry, University of California at Davis, One Shields Ave., Davis, CA 95616, USA }
	\author{Nicholas Curro}
	\affiliation{Department of Physics, University of California at Davis, One Shields Ave., Davis, CA 95616, USA}
	\author{Marina Radulaski}
	\email{mradulaski@ucdavis.edu}
	\affiliation{Department of Electrical and Computer Engineering, University of California at Davis, One Shields Ave., Davis, CA 95616, USA}

	% \fundinginfo{United States Department of Energy, Office of Basic Energy Sciences, Chemical Sciences, Geosciences and Biosciences Division, Grant/Award Number: DE-FG0205ER15693; }
	
	\begin{abstract}
		Color centers are versatile systems that generate quantum light, sense magnetic fields and produce spin-photon entanglement. We review how these properties have pushed the limits of fundamental knowledge in a variety of scientific disciplines, from rejecting local-realistic theories to sensing superconducting phase transitions. In the light of recent progress in material processing and device fabrication, we identify new opportunities for interdisciplinary fundamental discoveries in physics and geochemistry.
	\end{abstract}
	
	\maketitle
	
	\section{Introduction}
	
	At the turn of the century, the negatively charged nitrogen-vacancy (NV$^{-}$) center in diamond attracted great attention as a versatile system for quantum and nanoscale science and technology. Although the electronic properties and optical spectra of these point-defect ensembles have been well studied, breakthroughs in terms of control of solid-state spins were achieved only recently.  The technology has advanced to the point that magnetic resonance can be performed at the level of an individual color center, and via optical detection. From this work, the NV$^-$ center emerged as the most precise nanoscale magnetometer and this device is now making an impact on a variety of disciplines, from high-pressure superconductivity to electron-phonon processes in 2D materials. Joint advances in substrate processing and dynamical-decoupling techniques enabled a giant leap in NV$^-$ center spin-coherence time, now approaching a second \cite{Bar-Gill2013}. These long coherence properties enable new tests of quantum mechanics, including the first demonstration of loophole-free Bell inequality violation \cite{Hensen2015}.
	
	In the 2010s, researchers began looking into alternative color center systems and silicon carbide (SiC) emerged as a suitable wide-band-gap vacancy center host because of its intrinsically favorable coherence properties, high optical stability and ensemble emission homogeneity. Subsequently, the centro-symmetric class of diamond defects that incorporate group-IV atoms \deleted[id=MR]{(silicon-vacancy SiV$^-$, germanium-vacancy GeV$^-$, tin-vacancy SnV$^-$, lead-vacancy PbV$^-$)} was characterized to have narrowband optical emission and attractive coherence properties when cooled to low enough temperatures. All these color centers introduced new combinations of operational wavelengths, spin and orbital complexes, and have enabled new scientific explorations.
	
	This article focuses on the impact that color centers \added[id=MR]{in 3D semiconductors} have made on fundamental research across disciplines. First, we review the point-defect systems and associated techniques, then we survey the color center-enabled breakthroughs in a variety of fields, from skyrmions to Malaria studies, and, finally, we discuss opportunities for promising interdisciplinary discovery in areas of physics and geochemistry. The paper is intended to complement the insightful reviews on \added[id=MR]{the 3D, and recently 2D,} color-center applications to quantum information technologies \cite{atature2018material,awschalom2018quantum, doherty2013nitrogen} \deleted[id=VN]{and}\added[id=VN]{, applications to} photonic integration \cite{radulaski2019nanodiamond, bradac2019quantum} \added[id=VN]{, and engineering new color centers \cite{bassett2019quantum}}.

	\section{Color Centers}
	
	Color centers are light emitting defects in the lattice of semiconductors. The defects are formed by vacancies, misplaced atoms, foreign species atoms, or atom-vacancy complexes.

	\begin{figure}[ht]
		\centering
		\includegraphics[width=\linewidth]{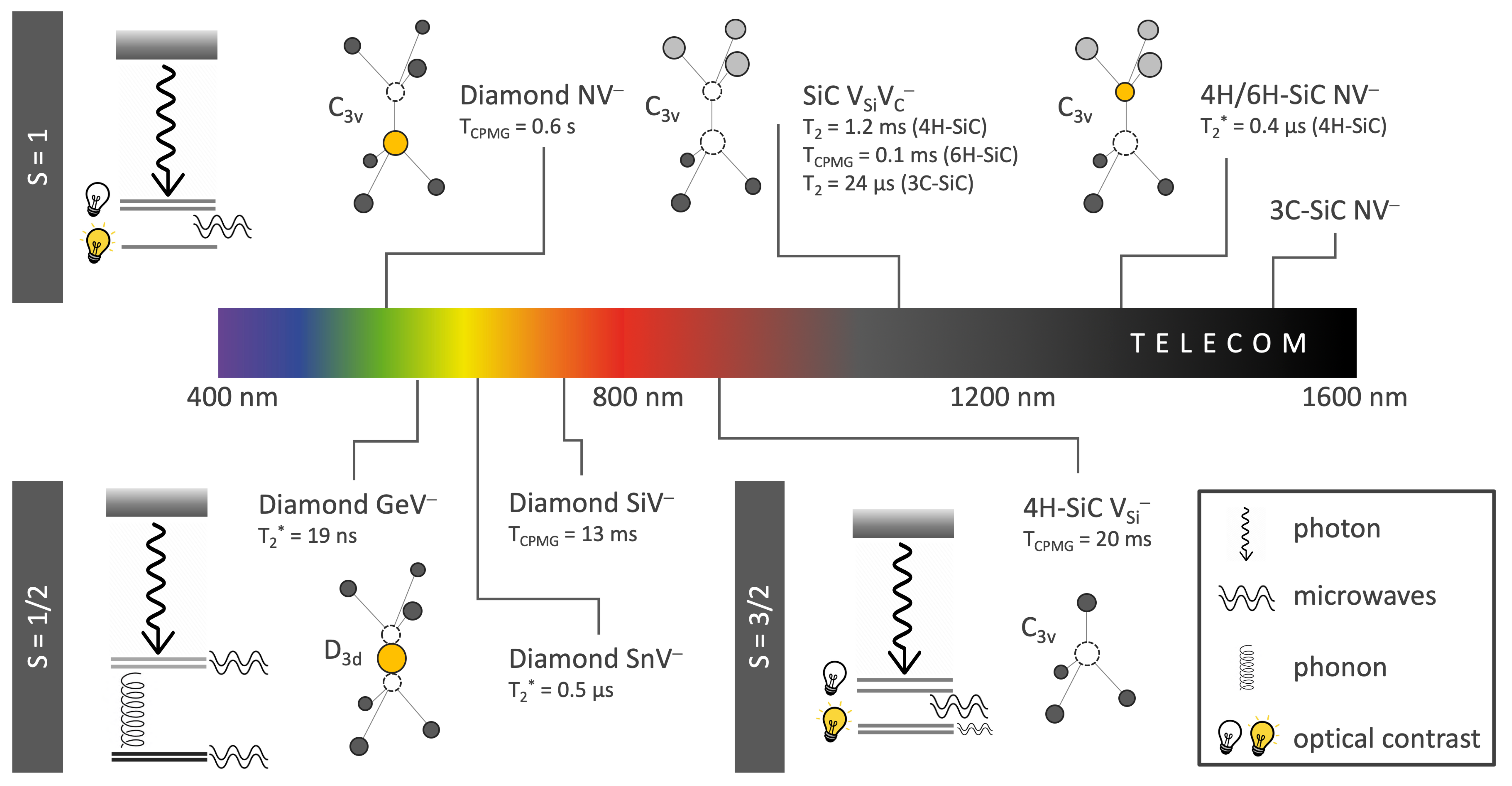}
		\caption{Prominently studied color centers in diamond and silicon carbide organized by the electron spin value which defines their energy scheme and associated interactions. The color centers are represented with their symmetry groups, crystal defect illustrations 
			\added[id=MR]{where yellow atoms represent foreign species and dashed outlines represent missing atoms}, emission wavelength on the spectrum and the longest measured coherence times\added[id=MR]{, where CPMG stands for the utilized Carr-Purcell-Meiboom-Gill dynamical decoupling sequence} \added[id=MR]{\cite{Bar-Gill2013, falk2013polytype, christle2015isolated, zargaleh2016evidence, von2016nv, simin2017locking, zargaleh2018nitrogen, mu2020coherent, siyushev2017optical, sukachev2017silicon, trusheim2020transform}}.}
		\label{fig:colorcenters}
	\end{figure}
	
\begin{figure}[ht]
		\centering
		\includegraphics[width=0.8\linewidth]{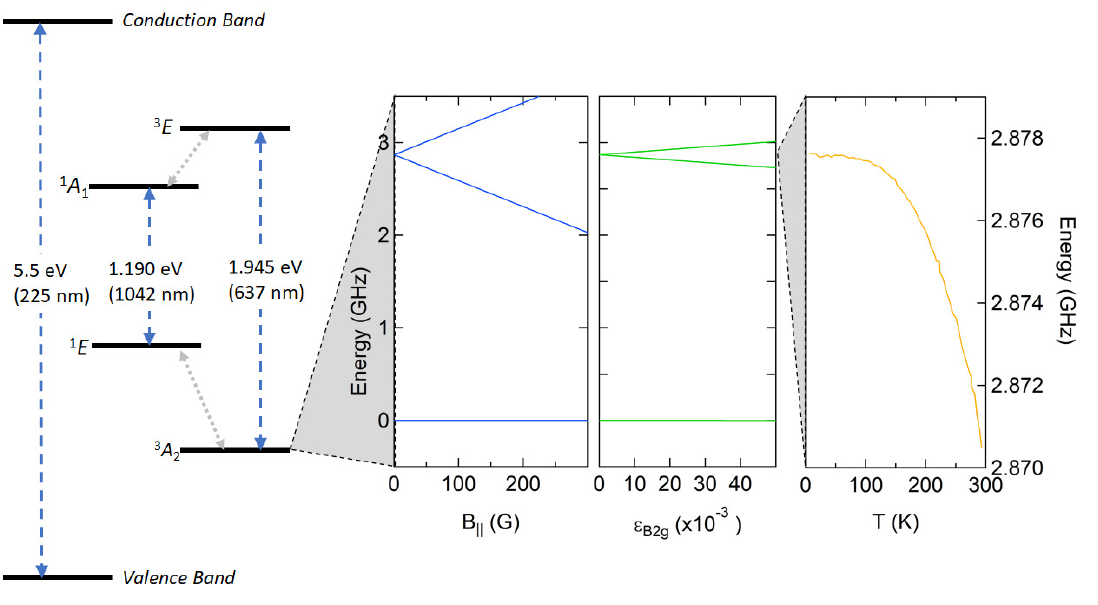}
		\caption{Energy levels of an NV$^{-}$ center  within the bandgap of the host diamond lattice.  From left to right: the metastable singlet ground and excited state, the ground and excited triplet states, the splitting of the $|0\rangle$ and $\pm |1\rangle$ states as a function of magnetic field (oriented parallel to the NV$^-$ axis) anisotropic in-plane strain \cite{Udvarhelyi2018}, and versus temperature \cite{chen2011temperature}. The temperature dependence has been magnified by a factor of 100 for clarity.}
		\label{fig:energylevels}
	\end{figure}

	A color center localizes electronic orbitals and behaves as a quasi-atom in a solid-state platform capable of emitting single-photons. Its symmetry and electron spin magnitude define a rich scheme of energy levels which can be coupled via dipolar (optical and microwave) and phonon interactions, and tuned via magnetic field, temperature, electric field and mechanical strain, as illustrated in Fig. \ref{fig:colorcenters} and detailed in Fig. \ref{fig:energylevels} for the NV$^{-}$ center in diamond.  In \deleted[id=VN]{this} \added[id=VN]{the diamond NV$^{-}$} case, the ground state is a spin triplet $S=1$, and dipole-dipole interactions between the electron spins lifts the degeneracy between the $|m_S=0\rangle$ and $|m_S=\pm 1\rangle$ states, giving rise to a \emph{zero field splitting}.  This degeneracy can be lifted further with magnetic field, and spectra can be obtained by measuring the optical emission intensity as a function of microwave frequencies (optically-detected magnetic resonance - ODMR).  Strain, electric field, and temperature also strongly affect the energy levels, which  lead to the enormous versatility of the NV$^-$ center as a sensor.

	\section{Techniques}
	A significant amount of effort has been put into developing the promising characteristics of color centers into usable tools \added[id=VN]{and techniques} 
	for research. This section focuses \added[id=VN]{primarily }on the developed \deleted[id=VN]{tools} \added[id=VN]{techniques} based on NV$^-$ centers, expanded by a discussion on the current state of the art for other color centers. Particular emphasis is given to the sensitivity and versatility of the described systems.
	
	\subsection{Quantum light emission and spin-photon entanglement}
	Interest in quantum light emission, detection, and manipulation is seen in many disciplines, from fundamental explorations of quantum mechanics to applications in quantum technologies. In color centers, the emission of single photons comes from the finite time it takes for a two-state system to become excited and then decay back into the initial state, a process illustrated in Fig. \ref{fig:energylevels} for diamond NV$^-$. This method of generating single photons is present and accessible at room temperature which makes color centers an attractive candidate for quantum communication.
	Single photon emission, as demonstrated by photon anti-bunching measurements, has been successfully established in diamond's NV$^-$  centers \cite{Kurtsiefer2000}, SiV$^-$ centers \cite{Neu2011}, GeV$^-$ centers \cite{iwasaki2015germanium}, SnV$^-$ centers \cite{trusheim2020transform}, PbV$^-$ centers \cite{trusheim2019lead} chromium defects \cite{Aharonovich2010}, as well as in silicon carbide's divacancies \cite{koehl2011room}, silicon vacancies \cite{Widmann2015},  carbon antisite-vacancy pairs \cite{Castelletto2014}, vanadium centers \cite{wolfowicz2019vanadium} and nitrogen-vacancy centers \cite{mu2020coherent}. Moreover, indistinguishable photons were characterized in diamond SiV$^-$ emission \cite{sipahigil2014indistinguishable}, furthering the application space to the entanglement distribution schemes.
	
	In addition to single-photon generation, some color centers also enable spin-photon entanglement generation. 
	Spin-photon entangled states between the electron spin ($\uparrow$, $\downarrow$) and the photon polarization ($\sigma_{\pm}$) give rise to the wavefunction:
	\begin{equation*}
	|\Psi\rangle = \frac{1}{\sqrt{2}}\left(|{\uparrow}\rangle|\sigma_{-}\rangle+|{\downarrow}\rangle|\sigma_{+}\rangle\right),   
	\end{equation*}
	allowing for long-range entanglement distribution that can be used, for example, \deleted[id=VN]{for} \added[id=VN]{in} Bell tests \cite{Togan2010, economou2016spin}.

	\subsection{DC magnetometry}
	
	\deleted[id=VN]{The NV$^-$}\added[id=MR]{Nitrogen-vacancy} centers have attracted broad interest as magnetic field sensors. \added[id=MR]{The optical ground state of the NV$^-$ center corresponds to the $S= 1$ spin triplet state.}  Optically-detected magnetic resonance (ODMR) measures the fluorescence intensity \added[id=MR]{contrast between $m_S=0$ and $m_S=\pm1$ substates} of either a single center or an ensemble in the presence of microwaves (MW) \deleted[id=MR]{that couple to the $S=1$ states} (see Fig. \ref{fig:energylevels}).  The most straightforward approach combines continuous-wave \added[id=MR]{MW drive} with fluorescence intensity measurements as a function of microwave frequency.  \deleted[id=VN]{If the frequency matches the zero-field splitting (ZFS) between the $m_S = 0$ ground state and the $m_S= \pm 1$ ground spin states, the fluorescence intensity will be suppressed} \added[id=VN]{Under a DC magnetic field, the optically detected resonance undergoes Zeeman splitting from which the magnitude of the magnetic field can be determined} \cite{Acosta2009}. \added[id=VN]{Due to the four crystallographic orientations of the NV$^-$ center, $\left( \left[1 \Bar{1}\Bar{1}\right] \left[111\right], \left[\Bar{1}\Bar{1}1\right], \textrm{and} \left[ \Bar{1}1\Bar{1} \right]\right)$, there can be up to 8 resonance peaks in the ODMR spectrum. The different splitting of these resonance peak pairs is dependent on the relative angle between the magnetic field and the corresponding orientation vector, which allows for the extraction of both magnetic field magnitude and direction \cite{Maertz2010}} \deleted[id=VN]{In the presence of a DC field, the $m_S=\pm1$ levels split and multiple resonances appear. Furthermore, the energy level structure depends critically on the angle between magnetic field, and the NV$^-$ axis. Because there are four equivalent orientations of the NV$^-$ axis in the diamond structure, a spectrum in a magnetic field exhibits multiple resonances.} The sensitivity of CW-ODMR has been shown to be $\sim10$ $\mathrm{nT/\sqrt{Hz}}$  at room temperature \cite{Acosta2010}, however, this approach suffers from power broadening \added[id=VN]{ which can be mitigated by pulsed magnetic resonance techniques}.

\begin{figure}
		\centering
		\includegraphics[width=\linewidth]{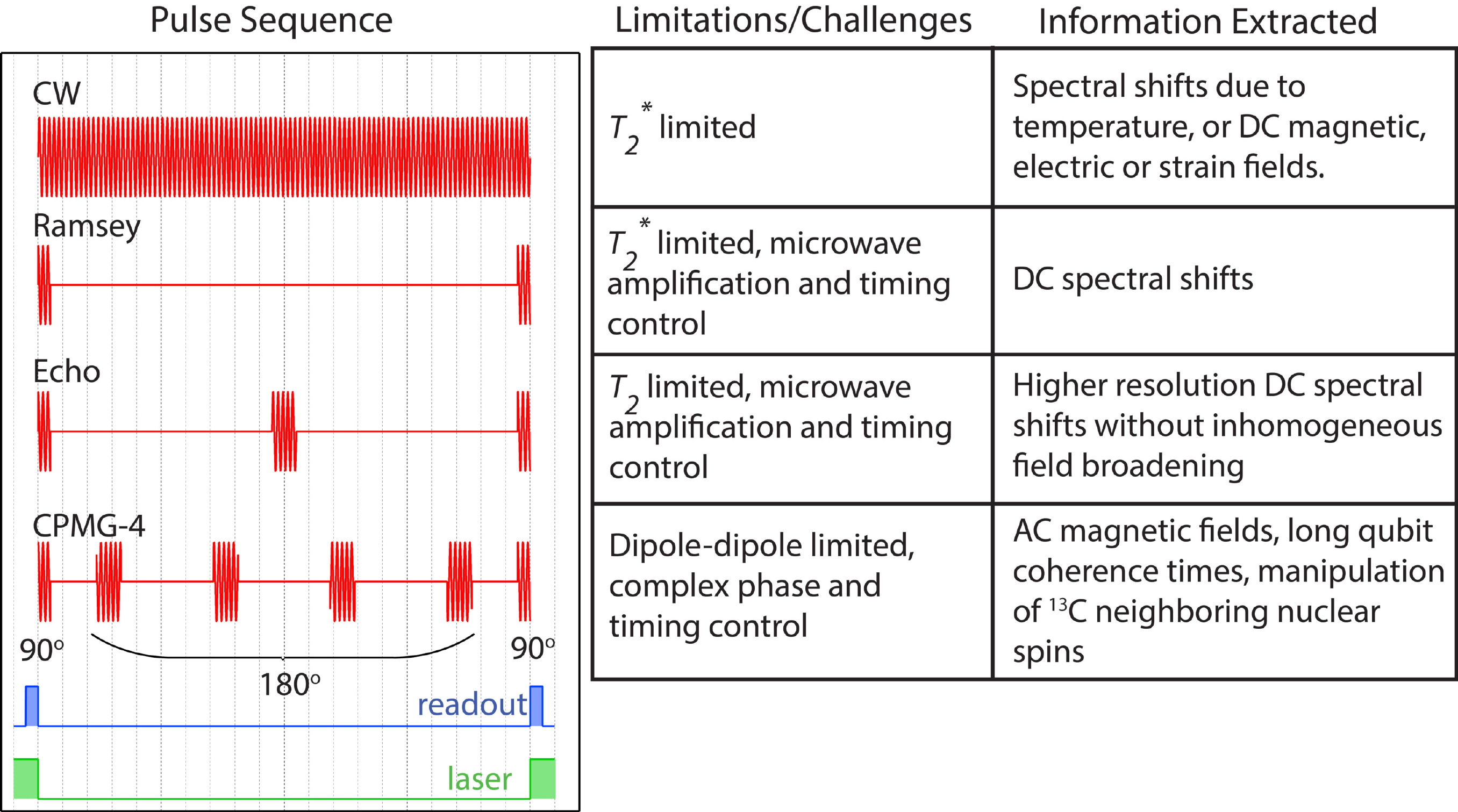}
		\caption{Comparison of different ODMR pulse sequences.  Initialization/Polarization is accomplished by laser illumination (green), and read out (blue) by comparing fluorescence intensity before and after the sequence. Microwave pulses (red) are introduced via an antenna located in the vicinity of the color centers.}
		\label{fig:pulsesequence}
	\end{figure}

	Pulsed magnetic resonance offers an improved approach, and utilizes both laser and microwave pulses \cite{Dreau2011}.  Laser pulses polarize the NV$^-$ spins to the $m_S=0$ state, and short microwave pulses manipulate the spin direction.  A Ramsey pulse sequence  (Fig. \ref{fig:pulsesequence}) is particularly sensitive to DC fields, but is limited by the field inhomogeneity through the dephasing time, $T_2^*$, and  requires sufficiently rapid Rabi oscillations that may not be viable for all samples \cite{barry2019}.   Dephasing can be improved through double quantum coherence (using two microwave transitions to achieve $\delta m_s = \pm2$ and overcome the limit of $\delta m_s = \pm1$)  \cite{bauch2018ultralong, Myers2017}, to decouple the NV$^-$ from spin noise in the diamond lattice \cite{bauch2018ultralong, knowles2014observing}.
	With state of the art readout techniques including near-infrared absorption \cite{Acosta2010}, efficient spin-to-charge conversion \cite{Shields2015} have led to measured DC magnetic field sensitivities as low as 15 $\mathrm{pT/\sqrt{Hz}}$ \cite{Chatzidrosos2017, Barry2016}.
	
	\subsection{Electric field sensing}
	
	The combination of high sensitivity and the wide range of temperatures over which NV$^-$ spectroscopy is effective makes color center electric field sensing a significant competitor in nanoscale electric field sensing technologies.
	The symmetry of the $^{3}A_2$ ground state of the NV$^-$ center gives rise to a linear Stark effect \cite{van1990electric,Tamarat2006}, which has been exploited to demonstrate ultra-sensitive electric field sensing under a wide range of conditions \cite{Dolde2011, dolde2014nanoscale, michl2019, Chen2017electrometry}. 
	\deleted[id=VN]{Because the Zeeman effect is stronger than the Stark effect in the ground state, s}\added[id=VN]{S}ensitive electric field measurements are made easier by a calibrated setup that aligns 
	a\added[id=VN]{n external} magnetic field in order to decouple the electric and magnetic field sensing capabilities of the NV$^-$ center. 
	\added[id=VN]{This is due to the significantly higher contribution of the Zeeman relative to the Stark effect in the ground state splitting. %: $\hbar\Delta\omega_\pm = d_{gs}^{\parallel}E_{z} \pm F(\Vec{B},\Vec{E}, \Vec{\sigma}) - F(\Vec{B}, \Vec{E}, \Vec{0})$ where (discarding higher order terms) $F(\Vec{B},\Vec{E}, \Vec{\sigma}) \propto \left(\mu_B^2g_e^2B_z^2 + d_{gs}^{\perp 2}\left(\Vec{E} + \Vec{\sigma}\right)_{\perp}^2\right)^{1/2}$. In the expected case, $d_{gs}(\Vec{E}+\Vec{\sigma})/\mu_Bg_eB <<1$. Therefore 
		Without careful alignment, to decouple the strong $B_z$ contribution from the electrical component, the splitting due to the electric field may become experimentally difficult to detect.} \cite{Dolde2011}. 
	Under ambient conditions a single NV$^-$ can sense  the change of a nearby defect from the neutral NV$^0$ state to the negatively charged NV$^-$, a change of only one electron \cite{dolde2014nanoscale}. 
	Additionally, with the use of a static bias voltage in place of the static magnetic field, NV$^-$ center ensembles have achieved an electric field sensitivity of $10^{-6}$ $\mathrm{V/\mu m}$\cite{michl2019,Chen2017electrometry}. 
	Using dynamical decoupling, NV$^-$ ensembles have shown sensitivity to AC electric fields down $10^{-7}$ $\mathrm{V/\mu m}$, which is similar to other solid-state electrometers at room temperature (see section 3.6.1).
	
	\subsection{Thermometry}
	
	Color center based thermometry enables high sensitivity temperature probes with sub-micrometer spatial resolution.  This capability, in turn, has the potential to explore a broad range of phenomena, including  processes at the subcellular scale in biota, molecular interactions, and micro-circuitry \added[id=VN]{\cite{Kucsko2013, laraoui2015imaging}}. The zero-field splitting (see Fig. \ref{fig:energylevels}) depends sensitively on temperature and is governed by local crystal lattice strains \cite{chen2011temperature}.
	In diamond, both NV$^-$ centers and, more recently, germanium-vacancy color centers, have proven to have good thermal sensitivity \deleted[id=VN]{\cite{doi:10.1021/acsphotonics.7b01465, neumann2013high}}.
	One of the major roadblocks to color center thermometry is that stray magnetic or strain fields can be mistaken for thermal effects when observing spectroscopic shifts of the triplet state. This misidentification will become even more important as more complex systems are probed. These challenges can be partially overcome by the application of a strong axial magnetic field \cite{neumann2013high}, coherent microwave pulsed control of spin-echo states \cite{Toyli2013}, and the use of isotopically pure $^{12}C$ diamond. The use of isotopically pure diamond reduces the interaction with the magnetic field fluctuation that arises from the $^{13}C$ nuclear spin bath. The use of NV$^-$ center thermometry is also limited by the high RF fields necessary to probe the NV$^-$ state, limiting sensitivity to $\sim$10 $\mathrm{mK/\sqrt{Hz}}$ \cite{neumann2013high, Toyli2013, PhysRevB.91.155404}. Despite these limitations, NV$^-$ thermometry has already been able to image sub-cellular length scales in human fibroblast using a combination of NV$^-$ enriched nanodiamonds and gold nanoparticles \deleted[id=VN]{\cite{Kucsko2013}}, and
	to image spatial variations of thermal conductivity at the nanoscale \deleted[id=VN]{\cite{laraoui2015imaging}}.
	Similar temperature-dependent zero-field splittings are also present for other color centers, such as the $S=3/2$ silicon vacancy in silicon carbide \cite{Kraus2014temp,Anisimov2016}.

	\subsection{AC magnetometry and NMR}
	Color centers are poised to revolutionize the fields of both nuclear magnetic resonance (NMR) spectroscopy and magnetic resonance imaging (MRI). \added[id=VN]{Since the advent of scanning NV magnetometry more than a decade ago, t}hese tools have become indispensable \deleted[id=VN]{tools} in diverse fields of research, including analytical chemistry, structural biology and physics \added[id=VN]{\cite{balasubramanian2008nanoscale,  taylor2008high}}. However, a major deficiency of conventional NMR is \deleted[id=VN]{the low sensitivity of coil-based induction, which} \added[id=VN]{that it} precludes NMR studies for samples with volumes below the microliter scale \added[id=VN]{because of the low-sensitivity of coil-based induction to small numbers of nuclear spins} \cite{2018natureWalsworth}. Field sensing by color centers offers an alternative method for NMR detection at the nanoliter scale.  The static field of a nucleus can create a dipolar field at the NV$^-$ center, but typically this field is on the order of pT to nT, which is too small to resolve spectrally. However, time-dependent AC detection is possible, in a manner similar to lock-in detection. Not only can NV$^-$ centers in diamond serve as single atom sensors at the nanometer length scale \cite{2012PRLLukin}, but they can detect the weak magnetic field fluctuations induced by an ensemble of statistically polarized nuclear spins in a nearby sample \cite{2013scienceRugar}. These properties have led to several recent developments of AC magnetometry using NV$^-$ sensors \cite{Bucher2019}.  Adaption of the NV$^-$ technology to solution NMR is already underway. The S=1/2 nuclei, such as $^1$H, $^{19}$F and $^{31}$P, have been detected from solutions that are on the diamond surface near the NV$^{-}$
	centers. 
	
	\subsubsection{Dynamical decoupling}
	
	The strong coupling between an NV$^-$ sensor and the local magnetic field is advantageous for detecting static magnetic fields, but is also a hindrance in the presence of noisy environments, such as the magnetic fields created by background $^{13}$C nuclear spins in the diamond lattice.  Random fluctuations of the local field, which can be on the order of  \added[id=NC]{$10^{-7}$T}, can quickly dephase the coherent precession of an NV$^-$ spin.   To remove the effect of the background noise, dynamical decoupling (DD) pulse sequences are often employed to continually refocus the precessing NV$^-$ spins (for example to form a Hahn echo)  (Fig. \ref{fig:pulsesequence} \cite{CPSbook}). This approach removes magnetic inhomogeneity and extends coherence lifetimes.  However, DD sequences can be designed with specific pulse timing so that an NV$^-$ center can be especially sensitive to fields at \emph{specific frequencies} that can be tailored to match the frequency of precession of nearby nuclei \cite{DDfilterPRL,DDfilterPRB,2015PRXCappellaro}. 
	By measuring the NV$^-$ response as a function of the DD filtering frequency, one can measure the NMR spectrum and identify the presence of multiple nuclear species. This approach has been demonstrated to detect nuclear spins located outside of the diamond lattice, where the field amplitude that can be  detected is of order 10 nT-rms \cite{2013scienceRugar,2012naturenanoWrachtrup,2014PRBPines,2015DuScience,2016ScienceLukin,2014naturecommJelezko,2014ScienceDegen}. 
	Single NV$^-$ centers offer the promise of combining AC sensing with high spatial resolution, especially because cantilevers with NV$^-$ centers within the tip are now commercially available  \cite{2008APLDegen,2015naturenanoAwschalom,Abobeih2019}.
	However, ensembles of NV$^-$ centers may result in a more subtle field sensitivity when more advanced data acquisition techniques have been utilized \cite{2015naturenanoWalsworthNMR}. Recent  work has focused on creating more sophisticated DD sequences that can suppress both disorder and interaction effects between NV$^-$ centers \cite{2019ArxivLukin}. Another important advance has been magnetic resonance imaging by NV$^-$ centers at the nanoscale by  combining DD sequences with pulsed field gradients \cite{2014naturecommCappellaro}.

	\subsubsection{Coherent averaging}
	
\begin{figure}
		\centering
		\includegraphics[width=\linewidth]{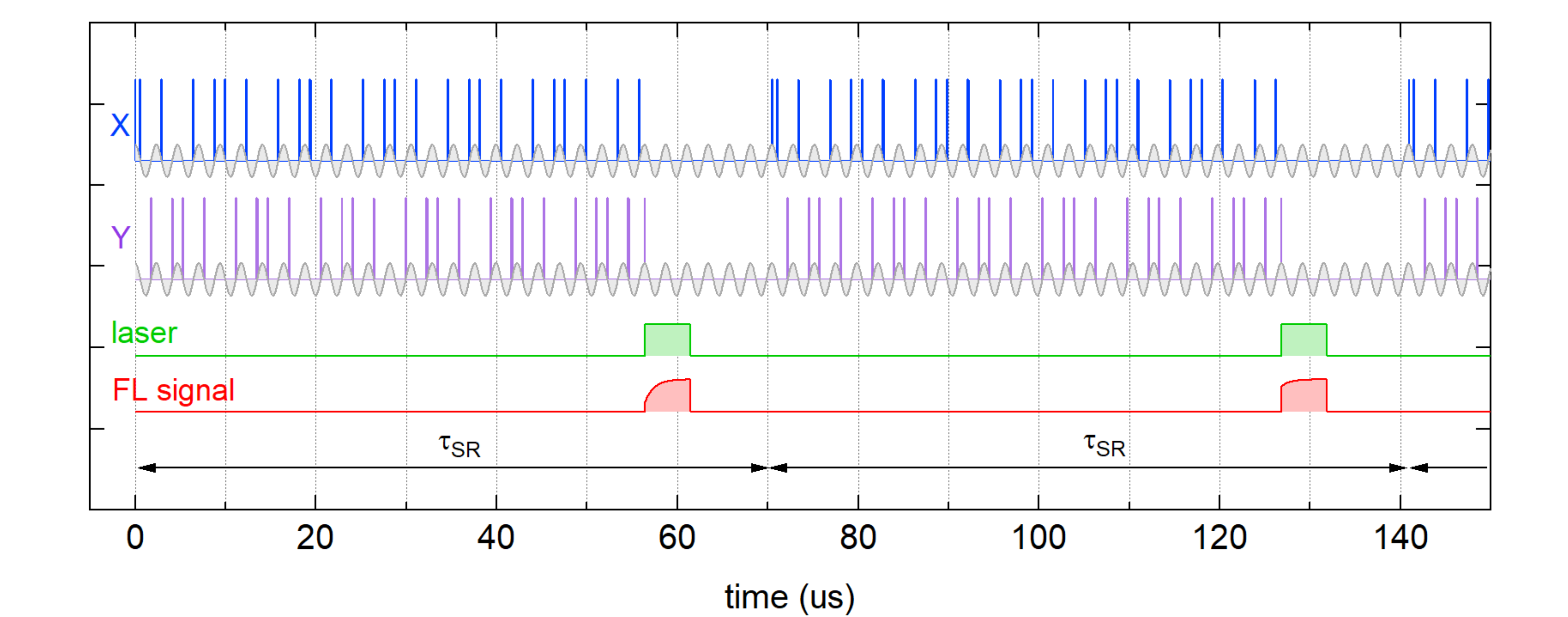}
		\caption{\emph{Quantum heterodyne} signal detection utilizing a dynamical decoupling pulse sequence at frequency $\tau_{SR}^{-1}$.  $X$ and $Y$ refer to the phase of the microwave pulses.  }
		\label{fig:DDnmr}
	\end{figure}

	Although pioneering experiments utilized the filter function of a DD sequence to map out the NMR spectra of different nuclear spin species on the surface of a diamond, the signal must be measured repeatedly at multiple frequencies and the spectral resolution is limited. A major advance has been the advent of coherent averaging of a continuously precessing nuclear spin field with multiple NV$^-$ detection sequences over the lifetime of the precessing nuclei, resulting in a dramatic enhancement of the NMR spectral resolution to below 1 mHz \cite{2018natureWalsworth, 2017ScienceJelezko}. 
	Since the $T_2$ of the nuclear magnetization is generally much longer than the time scale to measure the NV$^{-}$ centers with a DD sequence, one can then re-initialize the NV centers optically and repeat the measurement multiple times while the nuclear magnetization continues to precess (see Fig. \ref{fig:DDnmr}). The resulting NV$^-$ magnetization will be heterodyned with the DD frequency.   This \emph{quantum heterodyne} approach has  has been utilized to measure high resolution 1D NMR spectra of different molecules in solution at picoliter volume scales with sub-mHz resolution \cite{2018natureWalsworth, 2017ScienceJelezko, 2011natureOzeri,2017ScienceDegen}, and more recently multidimensional heteronuclear double resonance of organic molecules \cite{Smits2019}. At present the highest reported frequency resolution is $\sim400$ $\mu$Hz \cite{2017ScienceJelezko} and the highest  field sensitivity is around  $\sim30$ $\mathrm{pT/\sqrt{{Hz}}}$ \cite{2018natureWalsworth}. This is a transformative technology because it enables the determination of $J$ couplings in low fields on sub-nanoliter volume samples.  Furthermore, \deleted[id=NC]{the number of spins this method is capale of detecting is much smaller than the $10^{20}$ needed for conventional liquid NMR}\added[id=NC]{this approach can detect down to $10^{4}$ nuclear spins \cite{2017ScienceJelezko}, several orders of magnitude lower than the detection limit for conventional liquid NMR,} and solution volumes are sufficiently small that chemical shift resolution is sometimes limited by solute diffusion\deleted[id=NC]{, not pulsing}.
	
	\subsubsection{Hyperpolarization}
	A major challenge to AC magnetometry for NMR detection is achieving a sufficiently large nuclear magnetization.  Both single and ensemble NV$^-$ centers have similar field and frequency sensitivity, thus as long as an NV$^-$ ensemble is implanted in a shallow layer close to thermally polarized nuclei, nanoscale NMR is accessible \cite{2008APLDegen}. However, since ODMR is generally conducted at fields much lower than 1 T, the thermal polarization of the nuclei can lie below 10$^{-8}$ at room temperature so that the nuclear magnetization is quite small.  On the other hand, the polarization of the NV$^-$ spins can reach unity (hyperpolarization) by optical pumping.  An exciting possibility is to transfer some of the NV$^-$ polarization to the nuclear spins via cross-polarization using electron-nuclear double resonance techniques \cite{cai2013large}. \added[id=NC]{Increasing the nuclear polarization would increase the magnitude of the AC field created by the nuclei, and hence enhance the sensitivity of AC magnetometry, especially at lower fields.} To date, this technique can be utilized to create large hyperpolarizations of the $^{13}$C spins in the diamond lattice
	\cite{2015PRXCappellaro,2018ScienceAdPines,King2015, 2013Cai}. However, efforts to transfer this nuclear polarization \emph{outside the diamond lattice} have proved to be challenging\added[id=NC]{ \cite{NVnucleusHartmanHahnPRL,fernandez2018toward}}.  An exciting alternative method of
	hyperpolarization that has been reported is via transfer of polarization from
	electrons on nitrogen radicals in solution to adjacent  nuclear species \cite{bucher2018hyperpolarizationenhanced}. The signal enhancement is typically two to four orders of magnitude. \added[id=NC]{Yet another alternative employs sophisticated pulse sequences (via Hamiltonian engineering) to transfer polarization to the nuclear spins \cite{schwartz2018robust}.}

	\subsection{Fluorescent labelling}
	Because nanodiamonds (ND) are bio-compatible \cite{Vaijayanthimala_2009,Li2013,doi:10.1517/17425247.2015.992412} and have low cellular toxicity \cite{Mohan2010,Yu2005}, ND with NV$^-$ color centers has shown promise in biological applications, specifically in long term monitoring because of the photostability of the emission from the color center \cite{Yu2005,Fu2007}. Since, the NV$^-$ centers are inside the crystal lattice, the fluorescent properties of NV$^-$ centers are not influenced by surface modification of ND and the pH of the surrounding media \cite{10.1117/12.2004494}. The long fluorescence times of ND (20 ns) compared to that of cell and tissue autofluorescence (3 ns) can result in improved intracellular contrast \cite{10.1117/12.2004494}. Stimulated Emission Depletion (STED) microscopy was employed for superresolution cellular imaging using fluorescent ND \cite{doi:10.1002/anie.201007215}. The image resolution was also developed using techniques like two-photon microscopy \cite{Hui:10}. In the presence of strong background autofluorescence (arising from endogenous molecules), microwaves that are resonant with the crystal field splitting of ground state spin were applied and images with and without microwaves were subtracted to remove the autofluorescence (which does not change under the influence of microwave radiation) \cite{Igarashi2012}. These techniques highlight the versatility of the NV$^-$ rich nanodiamonds for future use in fluorescent tracking of tissues, individual cells and even intracellular processes.
	
	Silicon carbide is bio-compatible \cite{Oliveros2013,Bonaventura2019} and is known for its chemical inertness. Fluorescent 4H-SiC nanoparticles having excellent colloidal stability can be prepared using laser ablation \cite{Castelletto:17}. Thus there is a potential to extend the applications of SiC nano-particles for bio-imaging applications.

	\section{Fundamental findings}
	
\begin{figure}[ht]
		\centering
		\includegraphics[width=\linewidth]{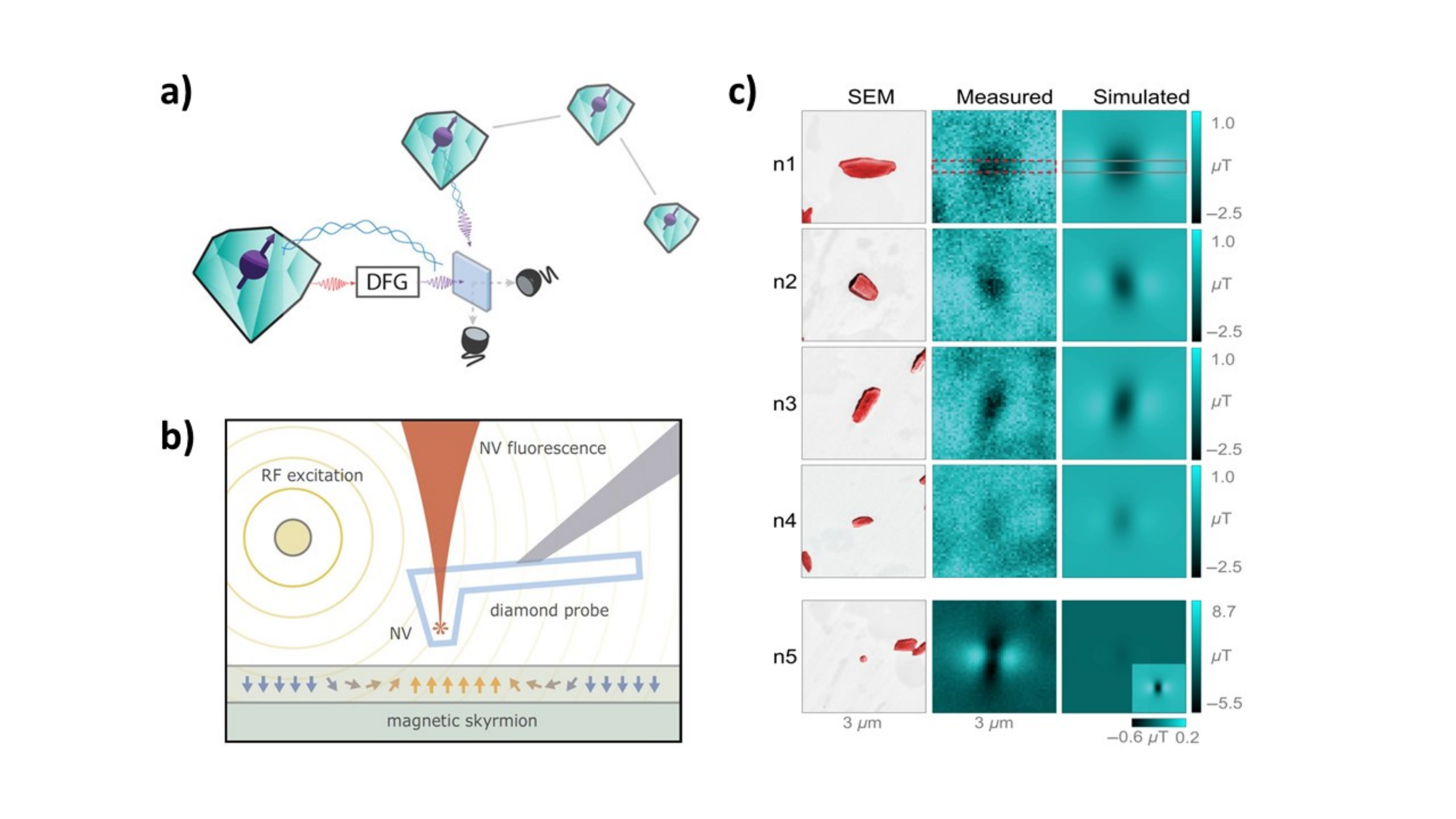}
		\caption{ a) A schematic of  photon-assisted entanglement distribution between distant diamond NV$^-$ spins, for Bell Inequality tests \cite{tchebotareva2019entanglement} (Reprinted with permission from Tchebotareva \textit{et al} \cite{tchebotareva2019entanglement}); b) Experimental setup for magnetic imaging of skyrmions using a single NV$^-$ center at the apex of a single crystal diamond probe \cite{PhysRevMaterials.3.083801} (Reprinted with permission from Jenkins \textit{et al} \cite{PhysRevMaterials.3.083801}); c) (from left to right) SEM images of individual natural hemozoin nanocrystals, their corresponding magnetic microscopy images, and simulated magnetic images used in studies of Malaria \cite{Fescenko2019} (Reprinted with permission from  Fescenko \textit{et al} \cite{Fescenko2019}).    
		}
		\label{fig:fundamentalfindings}
	\end{figure}

	From the quantum world of single photons to probing the biological matter, color centers provide robust systems for exploring fundamental scientific questions. In this section, we will highlight some experiments that have resulted in advances of basic scientific knowledge as shown in Fig. \ref{fig:fundamentalfindings}.
	
	\subsection{Bell inequality test}
	For more than 50 years, physicists have been \added[id=VN]{proposing and } performing experiments to challenge local realism and disprove the Einstein-Podolsky-Rosen (EPR) paradox
	\added[id=VN]{\cite{ClauserBell1969, Fry1976}}. 
	The combination of sophisticated optical measurement systems and color centers' single photon emission properties have provided a robust testing ground for Bell's Theorem and the existence of quantum entanglement.
	In 2015, experimenters successfully performed a Bell Test that closed three of the strong objections to previous tests. The detection loophole was closed by demonstrated efficient spin readouts of NV$^{-}$ centers. The locality loopholes were closed by the spatial separation of 1.3 km, far enough that the photons travelling at the speed of light could not communicate with each other before measurement, and fast random selection of measurement basis followed by the high-efficiency readout \cite{Hensen2015}.
	
	The limitations of the Bell Test setup described in Ref \cite{Hensen2015} notably include the rate of entangled photon production, a limitation that could be addressed in a number of ways discussed in this paper. Careful photonic engineering could increase the number of identically produced photons, which would increase the number of potential entanglement events per second. Another method of addressing this limitation is the use of a different color center that has a higher percentage of photons emitted into the zero-phonon line (ZPL) to increase the production rate of entangled photon pairs. Further explorations into extending the range of the light propagation has been performed by downcoverting emitted photons to telecommunication wavelengths \cite{tchebotareva2019entanglement}.
	
	\subsection{Probing magnetic properties of materials}
	\deleted[id=VN]{The sensitivity of color centers in both spatial and magnetic field regimes detailed above  has led to a number of discoveries in magnetism. Many interesting antiferromagnetic (AFM) systems are insulators, thus spin-polarized scanning tunneling microscopy (STM), the standard measurement system for nanoscale spatially-resolved low magnetic fields, can not be utilized. The NV$^-$ scanning offers a new approach of investigating complex AFM ordering. Researchers imaged nanoscale magnetic ordering in thin film BiFeO$_3$, a multiferroic material, using scanning NV$^-$ magnetometry, and contradicted previous conjecture by demonstrating the presence of the spin cycloid in thin films \cite{gross2017real}.}
	
	\deleted[id=VN]{The relative non-interacting nature of color center probes was} \added[id=VN]{The fact that NV magnetometers do not produce stray magnetic fields that will interfere with magnetic measurements, coupled with the excellent resolution they can achieve has led to the rise of NV magnetometry as one of the major techniques for measuring and imaging nanoscale magnetic features such as single skyrmions and AFM magnetic ordering, measurements that are experimentally difficult to achieve with other, more invasive techniques \cite{rondin2013stray}. It is possible to use conventional MFM in concert with NV$^-$s in diamonds and magnetic field simulations in order to help separate out the magnetic fields generated by the MFM probe; in one case, nanodiamonds were scattered across the surface of FeGe in order } \deleted[id=VN]{ used in concert with magnetic force microscopy and simulations}to investigate the movements of domain wall structures in the helimagnet\added[id=VN]{ic phase of} FeGe. \deleted[id=VN]{The scanning NV$^-$ system used in this result was of } \added[id=VN]{This result was of} particular impact because the researchers were able to demonstrate that the domain wall motion was a product of the material itself, and not due to effects from the probe, a statement they would have been unable to make with conventional magnetic force microscopy or STM \cite{Dussaux2016}.

	\added[id=VN]{Even more interestingly than being a complementary technique to other methods of magnetometry, NV magnetometry has been increasingly used as a standalone technique for complex magnetic systems. For example,  many interesting antiferromagnetic (AFM) systems are insulators, thus spin-polarized scanning tunneling microscopy (STM), the standard measurement system for nanoscale spatially-resolved low magnetic fields, can not be utilized. The NV$^-$ scanning offers a new approach of investigating complex AFM ordering. Researchers imaged nanoscale magnetic ordering in thin film BiFeO$_3$, a multiferroic material, using scanning NV$^-$ magnetometry, and contradicted previous conjecture by demonstrating the presence of the spin cycloid in thin films \cite{gross2017real}. }
	
	The discovery of room temperature stable skrymionic materials has led to a need for room temperature nanoscopic imaging, which is difficult to do with most conventional methods. Researchers explored magnetostatic twists in room-temperature skyrmions via nitrogen-vacancy center spin texture reconstruction. They found a Neel-Type skyrmion in Pt/Co/Ta that is \emph{not} left-handed, contrary to other Dzyaloshinskii Moriya interation-based reports in similar materials. A consequence of these new results is proposed tube-like skyrmion structures for which the chirality varies according to film thickness \cite{dovzhenko2018magnetostatic}. 
	
	Another skyrmion measurement in Ta/CoFeB/MgO probed the dynamics of skyrmion bubbles. The sizes of skyrmion bubbles are determined dynamically via domain wall hopping between vortex pinning sites. These domain walls exhibit significant increases in magnetic fluctuations. The imaging showcases the important relationship in the interactions between internal degrees of freedom in bubble domain walls and pinning sites in thin films and provides direct measurements of how skyrmion movement deviates from micromagnetic predictions when the length scales of material defects are smaller than skyrmion size \cite{PhysRevMaterials.3.083801}. In a different experiment,
	
	The wide range of temperatures at which color centers work can provide an important insight into temperature dependent dynamics of materials. One of the major avenues of research opened up by this ability to measure small magnetic fields at a wide range of temperatures is new measurements of magnetic dynamics within superconductors. Researchers measured magnetic field expulsion in various types superconductors including standard type-I, conventional type-II, and several iron pnictide samples. The category of superconductor was found to heavily influence the method of and extent to which Meissner explusion occurs when cooled in a static magnetic field around $T_c$ \cite{nusran2018spatially}. As color center magnetometry works over large magnetic field ranges, single NV$^{-}$ centers have been used to explore superconductor create real-space maps of vortices \cite{pelliccione2016scanned} and to observe vortex pinning of individual vortices both using wide field imaging \cite{schlussel2018wide} and scanning methods \cite{thiel2016quantitative}. In a notable case, pinning measurements of Pearl vortices diverged from the commonly used monopole approximation of such objects \cite{thiel2016quantitative}.
	
	\subsection{2D material measurements}
	The high spatial resolution of NV$^-$ center measurements makes these probes excellent candidates for exploring local electrical properties in 2D materials. Of particular interest are the non-equilibrium dynamics in gated devices that provide insight into the fundamental properties of 2D materials like graphene. Probing graphene properties via NV$^-$ centers has particularly piqued the interest of many researchers who have measured current flow near defects \cite{Tetiennee1602429}. Spatially-resolved, room temperature NV-magnetometry has been used to measure viscous Dirac fluid flow in graphene, which supports the theory that graphene acts as a Dirac fluid even under ambient conditions \cite{ku2019imaging}. Recently, evidence for the electron-phonon Cerenkov instability was found in graphene by measuring noise through spatially-resolved spin-relaxation rate \cite{andersen2019electron}. Finally, as researchers have started to design more complex systems (e.g. graphene field effect transistors), potentially interfering interactions between patterned graphene nanoribbons and the diamond substrate due to the applied gate voltage have been characterized. These interactions are currently being investigated and several solutions have been proposed to minimize their effects \cite{lillie2019imaging}.
	
	\added[id=VN]{Scanning NV probes have also played a significant role in exploring local magnetic properties in 2D materials. Researchers were able to measure atomically thin layers of the 2D vdW Magnet, CrI$_3$ and determine that magnetic properties were 'switched` on and off depending on whether the number of layers in the sample were odd or even\cite{thiel2019probing}. This kind of measurement is yet another illustration of how powerful scanning NV probes are when compared to other, more invasive, methods that are limited by micron-scale resolution or that generate their own magnetic traces that interfere with measurements.}
	
	\subsection{Biological applications}
	In the biological sciences, researchers have been using the bio-inert properties of diamonds and the high sensitivity of color centers to make great strides in a wide range of fields.
	
	Hemozoin crystals are biomarkers for malaria that exhibit paramagnetic properties. Spatially resolved magnetic characterization of these crystals is of particular interest because several antimalarial drugs target hemozoin crystal formation. The sensitivity of NV$^{-}$ center ODMR allowed the researchers to measure magnetic properties of individual nanocrystals. Superparamagnetism was found in a small, but significant sample of both natural and synthetic hemozoin crystals \cite{Fescenko2019}.
	
	In vivo measurements are already being advanced by color center sensing, with extreme demonstrations of sensitivity like of sub-cellular thermometry in human cells \cite{Kucsko2013} and the measurement of neuonal action potentials in an alive worm \cite{Barry2016}. As an example, fluorescent nanodiamonds (FNDs) are easily taken up by cells, but rarely expelled. FNDs have been used to label and track in vivo lung stem cells from transplantation to engraftment in mice in both healthy and damaged lungs over a week. This tracking enabled the quantitative analysis of where transplanted stem cells engraft in damaged lungs \cite{wu2013tracking}. 
	
	\section{Future fundamental explorations}
	\added[id=VN]{With the proven capability to push new fundamental findings demonstrated thoroughly in the above sections, and the rise of commercially available NV-based measurement systems through startup companies like Qzabre and Qnami, the era of color center-based based scientific exploration is just beginning.} \deleted[id=VN]{Development of color center systems enable tests of new fundamental questions across disciplines.} In this section, we will present the state-of-the art in color center platform development, and follow with ideas for their utilization for \deleted[id=VN]{the} novel fundamental discovery.

	\subsection{Material processing and photonic device integration}
	
	The advances in photonics and material processing are making leaps in color center performance, including the defect generation precision, coherence time, and spin state readout fidelity.
	
	\subsubsection{Isotopic purification}
	
	The naturally existing isotopic elements that possess nuclear spin ($^{13}$C, $^{29}$Si, $^{30}$Si) interact with the color center electronic spin, thus having a direct impact on its coherence. The coherence time can be improved by isotopic purification i.e., using isotopes that have zero nuclear spin during the material growth. For example, isotopically purified diamond was grown using isotopically enriched CH$_{4}$ in plasma-enhanced chemical vapor deposition (PECVD) \cite{Ohno_2012} and microwave plasma chemical vapor deposition (MPCVD) \cite{doi:10.1021/nl402286v} techniques. In the case of 4H-SiC, isotopically purified material was grown by sublimating a $^{28}$Si enriched polycrystalline SiC \cite{KARPOV2000347,PhysRevX.6.031014}, while for 3C-SiC, the molecular beam epitaxy and mass separation of ions was employed to grow isotopically purified 3C-SiC on Si substrate  \cite{HORINO1995657,doi:10.1002/1521-396X(200201)189:1<169::AID-PSSA169>3.0.CO;2-6}.
	
	\subsubsection{Generation of color centers}
	Color centers are defects in the lattice of a crystalline material that exist naturally in an as-grown material or can be doped \emph{in situ} during the growth of the material or can be implanted into the lattice. To implant the defects, projectiles (electrons, protons, neutrons, ions) with sufficient energy irradiate material surface and their interaction with the lattice generate defects in the material. The implantation is followed by annealing in vacuum or inert atmosphere to heal the damaged material and activate the implants. Since the ions and vacancies in the lattice diffuse at annealing temperatures, the temperature is optimized to control the generation of color centers. In case of SiC annealing, a graphite cap layer is sometimes used to prevent Si evaporation from the material during the anneal.
	
	Lattice strain affects optical properties of color centers. To circumvent strain issues, novel methods of material growth have been developed. Examples include the introduction of Si atoms during MPCVD growth of diamond \cite{PhysRevB.89.235101} for SiV$^-$ generation, shallow ion Sn implantation and diamond overgrowth \cite{rugar2019generation} for SnV$^-$ generation, vanadium impurity incorporation during the epitaxial growth of 4H-SiC using vanadium tetrachloride \cite{PhysRevApplied.12.014015}.
	
	A good spatial positioning of the color centers is important for better integration with photonic structures, but the implantation and irradiation methods lack such a control. To improve lateral positioning, lithographically patterned resist has been used to implant SiC with NV$^-$ centers \cite{wang2019coherent} and silicon vacancies \cite{PhysRevApplied.7.064021} at defined positions. Focused ion beam (FIB) has also been employed to implant silicon vacancies in diamond nanostructures \cite{Schroder2017} and 4H-SiC with tens of nanometers of position accuracy \cite{Wang2017}. Yields of the defect generation were improved using femtosecond laser writing to create silicon vacancy centers in 4H-SiC  (up to 30\%) \cite{Chen2019} and NV$^-$ in diamond (near-unity yield) \cite{Chen:19,Chen2017}.

	\subsubsection{Emerging color centers}
	
	The developing material processing approaches, as well as theoretical predictions, have enabled a new generation of color centers to be characterized. These emerging quantum emitters expand the diversity of spin complexes and emission wavelengths in the field, as shown in Table \ref{tab:1}. Among them are color centers emitting at telecommunication wavelengths (vanadium and NV$^-$ centers in SiC) suitable for optimized fiber-propagation, as well as those with desirable spin-photon entangling interfaces (NV$^-$center in SiC \cite{economou2016spin}).
	
	\begin{table}[tb]
		\centering
		%\begin{fourparttable}
		\begin{tabular}{llll}
			
			Substrate & Defect & Spin & Wavelength\\
			\hline
			4H-SiC & chromium (Cr$^{4+}$) impurity \cite{koehl2017resonant, Diler2020} & S = 1 & 1042-1070 nm \\
			4H-SiC & molybdenum (Mo$^{5+}$) impurity \cite{bosma2018identification} & S = 1/2 (or 1) & 1076 nm\\
			4H-SiC & nitrogen-vacancy \cite{zargaleh2016evidence} & S = 1 & 1252-1291 nm\\
			4H-SiC & vanadium (V$^{4+}$) impurity \cite{spindlberger2019optical} & S = 1/2 & 1280-1333 nm \\
			6H-SiC & molybdenum impurity \cite{bosma2018identification} & S = 1/2 & 1121 nm\\
			6H-SiC & nitrogen-vacancy \cite{von2016nv} & S = 1 & 1240-1305 nm\\
			6H-SiC & vanadium (V$^{4+}$) impurity \cite{schneider1990infrared} & S = 1/2 & 1309-1388 nm\\
			3C-SiC & nitrogen-vacancy \cite{zargaleh2018nitrogen} & S = 1 & 1468 nm\\
			diamond & lead-vacancy \cite{trusheim2019lead} & S = 1/2 & 520 nm\\
			diamond & neutral silicon-vacancy \cite{Rose60} & S = 1 & 946 nm\\
			\hline 
			\\
		\end{tabular}
		%\end{fourparttable}
		\caption{Properties of the emerging color centers in diamond and silicon carbide.\label{tab:1}}
	\end{table}

	\subsubsection{Passive photonic devices}
	
\begin{figure}
		\includegraphics[width=.8\linewidth]{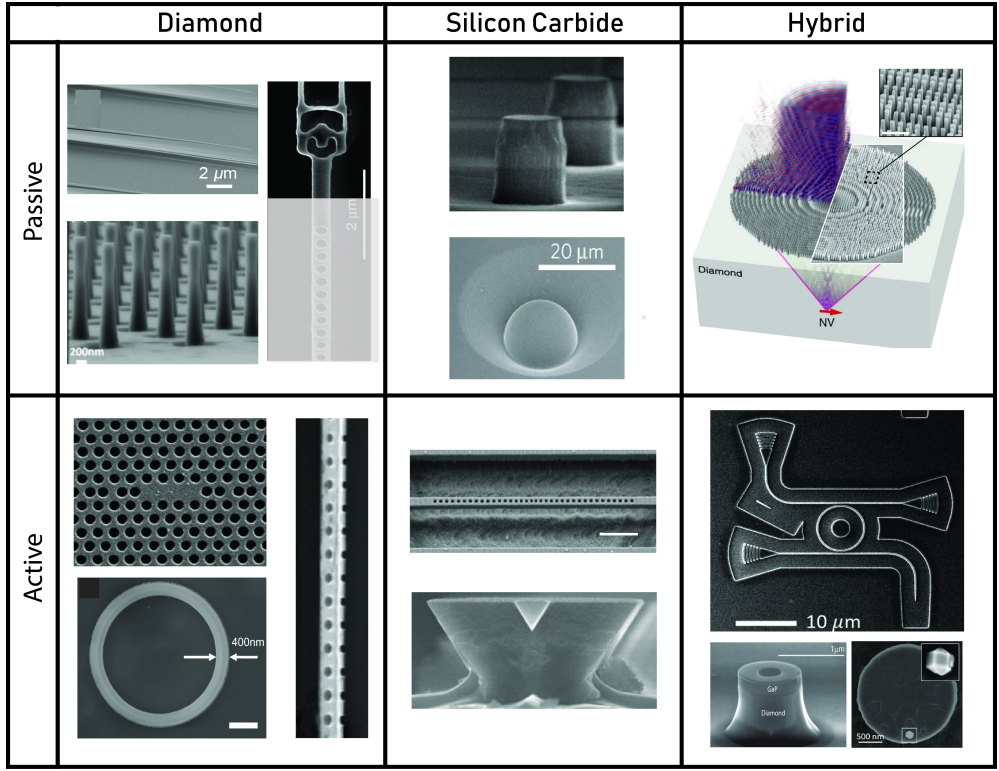}
		\caption{The SEM images of (clockwise from top left) - \textbf{Passive devices in diamond:} Free standing diamond tapered waveguides for efficient coupling \cite{PhysRevApplied.8.024026} (Reprinted with permission from Burek \textit{et al} \cite{PhysRevApplied.8.024026}), Inverse designed vertical coupler in diamond \cite{Dory2019} (Reprinted with permission from Dory \textit{et al} \cite{Dory2019}), Array of diamond nanowires \cite{Hausmann_2011} (Reprinted with permission from Hausmann \textit{et al} \cite{Hausmann_2011}); \textbf{Active devices in diamond:} 2D crystal cavities with defects (L3 cavity) \cite{doi:10.1063/1.5021349} (Reprinted with permission from Wan \textit{et al} \cite{doi:10.1063/1.5021349}), 1D nanobeam photonic cavities fabricated using angled-etching \cite{sun2018cavity} (Reprinted with permission from Sun \textit{et al} \cite{sun2018cavity}), Microring resonator in (111) diamond membrane on top of SiO$_2$ substrate, containing SiV \cite{regan2019photonic} (Reprinted with permission from Regan \textit{et al} \cite{regan2019photonic}); \textbf{Passive devices in SiC:} Nanopillars in 4H-SiC \cite{radulaski2017scalable} (Reprinted with permission from Radulaski \textit{et al} \cite{radulaski2017scalable}), Solid immersion lens fabricated on the surface of 4H-SiC \cite{Widmann2015} (Reprinted with permission from Widmann \textit{et al} \cite{Widmann2015}); \textbf{Active devices in SiC:} 1D nanobeam photonic crystal cavity in 4H-SiC \cite{Bracher4060} (Reprinted with permission from Bracher \textit{et al} \cite{Bracher4060}), 1D nanobeam with triangular cross-section fabricated using oblique plasma etching in 4H-SiC \cite{doi:10.1063/1.5058194} (Reprinted with permission from Song \textit{et al} \cite{doi:10.1063/1.5058194}); \textbf{Hybrid passive devices:} Pillars on the surface of single crystal diamond to create an immersion metalens \cite{Huang2019} (Reprinted with permission from Huang \textit{et al} \cite{Huang2019}); \textbf{Hybrid active devices:} Si ring resonator on 4H-SiC \cite{doi:10.1063/1.5116201} (Reprinted with permission from Wang \textit{et al} \cite{doi:10.1063/1.5116201}), Hybrid SiC-nanodiamond microdisk array \cite{radulaski2019nanodiamond} (Reprinted with permission from Radulaski \textit{et al} \cite{radulaski2019nanodiamond}), Hybrid GaP-diamond whispering gallery mode nanocavity \cite{PhysRevX.1.011007} (Reprinted with permission from Barclay \textit{et al} \cite{PhysRevX.1.011007}).}
		\label{fig:devices}
	\end{figure}

	Integration of a color center into a photonic device couples its emission with a designed electromagnetic mode. Depending on the light-matter interaction intensity, the devices can be divided into passive and active, as showcased in Figure \ref{fig:devices}. \added[id=SM]{Passive devices change the direction of the emitted photons and improve the light coupling efficiency into an optical mode of a device whereas active devices, in addition, also change the rate of emission of a color center.} \deleted[id=SM]{Passive photonic devices are used for improving the light coupling efficiency.} Often the desired coupling \added[id=SM]{from a passive device} is into an objective lens or a fiber, where the number of collected photons defines the bandwidth of the quantum channel or the sensitivity of the optical spin readout.
	
	Light generated by color centers in high refractive index bulk experiences refraction at the substrate-air interface, thus limiting the light collection into an objective lens. Photonic devices such as solid immersion lenses (SILs) \cite{Widmann2015, doi:10.1063/1.3519847, doi:10.1063/1.3519849} and, the more scalable, nanopillars \cite{radulaski2017scalable, Lukin2019, Hausmann_2011, zhang2015hybrid} have been used to increase the collection efficiency several-fold, while preserving the optical and spin properties of color centers in diamond and SiC.

	Similar advances have been achieved in hybrid platforms, where the ease of the fabrication of traditional photonic substrates was combined with the harder-to-etch color center hosts. Thereby, nanodiamonds containing silicon vacancy were placed on top of 3C- and 4H-SiC pillars \cite{zhang2015hybrid} and gallium nitride metalens were fabricated on top of the NV$^-$ center in bulk diamond efficiently coupling its emission into low-NA optics \cite{Huang2019}.
	
	Finally, to increase the light coupling in quantum optical circuits, diamond waveguide coupler design has been optimized. The inversely designed free-space grating couplers reached the efficiency of 25\% \cite{Dory2019}, while the tapered waveguide fiber couplers demonstrated $>$90\% efficiency \cite{PhysRevApplied.8.024026} and was successfully used to demonstrate an integrated color center quantum register \cite{nguyen2019integrated}.

	\subsubsection{Active photonic devices}
	Photonic cavities confine light to the (sub)wavelength volumes, $V \lesssim (\lambda/n)^3$. The resulting light-matter interaction can be used to modify the density of states in an embedded color center, thus enhancing its emission rate via Purcell effect. This method is often explored to boost emission branching into the ZPL, where indistinguishable photons are created. The enhancement of light-matter interaction is quantified by the Purcell factor $F$ which is directly proportional to the quality factor of the photonic cavity ($Q$) and inversely proportional to the mode volume ($V$) 
	\cite{Santori_2010}.
	
	Different geometries of photonic cavities, such as the disk shaped resonators that confine whispering gallery modes and photonic crystal cavities (PCC) that use distributed Bragg reflection to localize light to a small region, have been demonstrated. The resonant frequency is fine tuned via geometric parameters  (e.g. disk diameter, lattice constant, sizes of holes) in order to match the color center emission. Single defects were successfully integrated into photonic cavities either through the implantation/irradiation prior to the device realization or through targeted implantation into the fabricated cavity region.
	
	1D nanobeam PCC, 2D PCC and microring resonator fabricated in diamond, coupled to NV$^-$ centers \cite{Hausmann2013,doi:10.1063/1.4904909,Li2015,PhysRevLett.109.033604,Faraon2011} and silicon vacancies \cite{Zhang2018,PhysRevB.100.165428,riedrich2012one} enhanced emissions up to 70 times. SiC 1D nanobeam PCC and 2D PCC were coupled to silicon vacancy \cite{Lukin2019,Bracher4060}, \added[id=SM]{neutral divacancy \cite{crook2020purcell}} and Ky5 color centers \cite{doi:10.1063/1.4890083,PhysRevApplied.6.014019} had up to 120 times enhancement. To improve vertical confinement and collection efficiencies, angled-etching technique was used to fabricate nanobeam cavities with triangular cross-sections having high Q-factors in diamond \cite{Burek2014,doi:10.1063/1.4982603} and 4H-SiC \cite{doi:10.1063/1.5058194}. Other geometries like 2D PCC \cite{Song:11, Yamada:12, radulaski2013photonic, Song:19}, microdisk resonators \cite{zhang2015hybrid,Lu:13,doi:10.1063/1.4863932} were demonstrated in SiC.
	
	Hybrid devices are made with material of high refractive index on top of the material with color centers and the color center is aligned with the polarization vector of the evanescent fields of the photonic cavity \cite{Santori_2010}. GaP on diamond hybrid nanocavities and microdisk resonators coupled to NV$^-$ centers in diamond \cite{PhysRevX.1.011007,Santori_2010}, nanodiamonds containing silicon vacancy and Cr-centers on top of 3C-SiC microdisks \cite{radulaski2019nanodiamond} showed enhancement of emission by up to 17 times. Silicon ring resonator on top of 4H-SiC had an estimated F up to 36 for perfect dipole alignment and divacancy emitter located below the surface \cite{doi:10.1063/1.5116201}.

	The simulated Q-factors for each of these structures are not realised in the fabricated devices due to various factors like imperfections in the fabricated structures, lack of control over the precise position of the color center, losses in the material, surface roughness effects etc. Surface passivation of 4H-SiC surface using AlN improved the photostability, emission and charge state stability of carbon antisite-vacancy color center \cite{doi:10.1002/adma.201704543}. For hybrid devices, the achievable Q/V is limited by radiative losses into the substrate and this could be mitigated by extending the waveguide layer pattering into the substrate \cite{Santori_2010}.
	
	\subsubsection{\added[id=SM]{Opto-electronic devices}}
	\added[id=SM]{In addition to photonic devices, recent work on opto-electronic devices has expanded the toolset for controlling the emitter charge state, optical excitation and spin readout. Pioneering work has been done to integrate color centers in diamond \cite{doi:10.1063/1.4919388, Fedyanin_2016,  PhysRevLett.118.037601, Siyushev728} and 4H-SiC \cite{doi:10.1063/1.5004174, doi:10.1063/1.5032291, Niethammer2019, Anderson1225, Widmann2019} into semiconductor devices like Schottky diode, p-i-n diodes and light emitting diodes which may simplify experiments by reducing the complex optical control to integrated electronic circuitry. Hereby, electrical excitation was used to trigger single photon emission or electrodes were used to readout the charge state of color centers enabled by spin-selective photo-ionization.  Dynamic control of emitter charge states has also been demonstrated, providing an on-off knob to switch an experiment that depends on the reliable state of a color center including spin polarizability and ability of optical spin-state readout. Finally, electric field has been applied to tune the emission wavelength of SiC divacancies, which can be used to produce indistinguishable photons from different color centers and distribute entanglement between their spins.}

	\subsection{Fundamental experiments}
	
\begin{figure}
		\centering
		\includegraphics[width=0.5\linewidth]{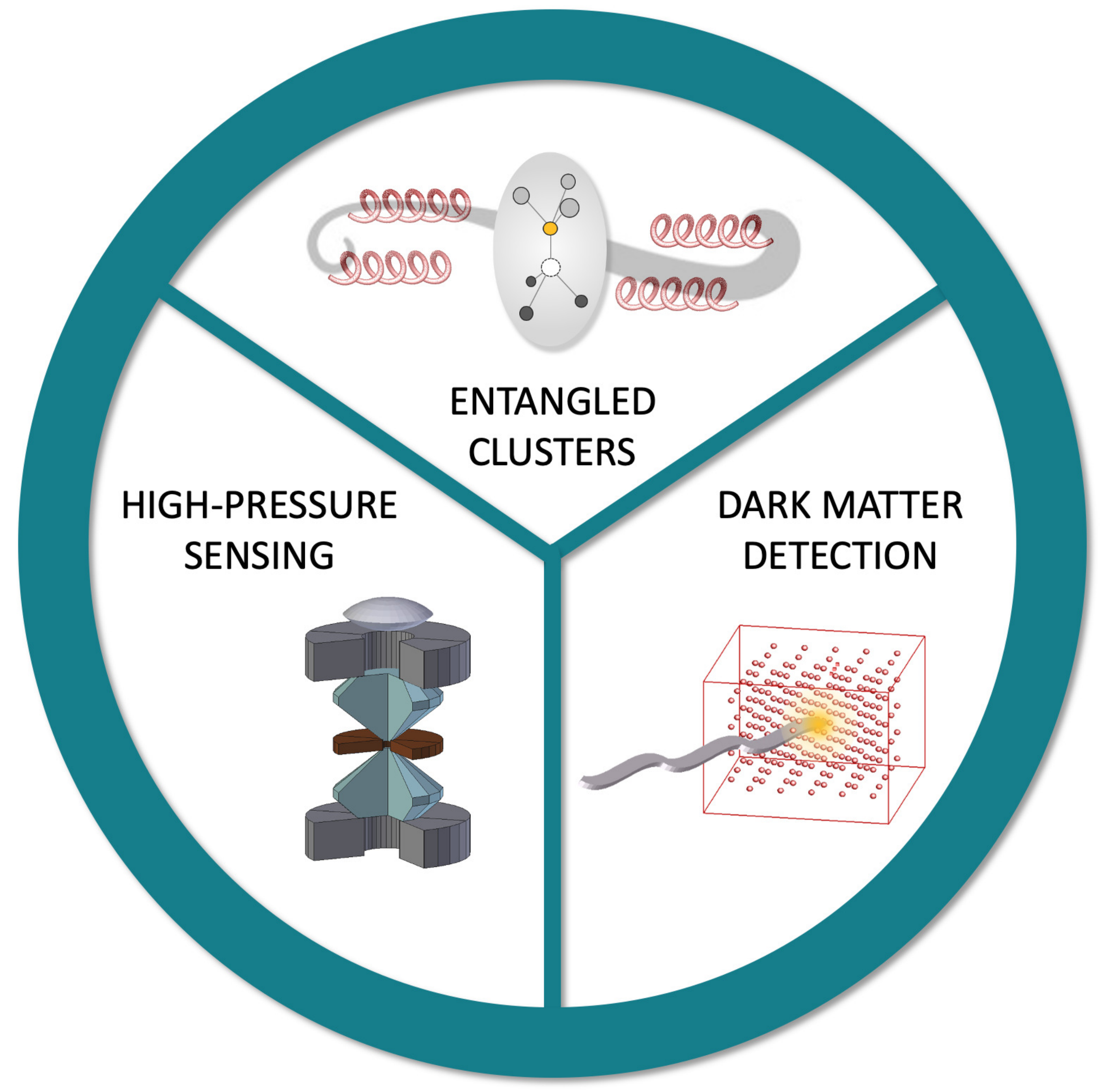}
		\caption{Color centers have a promising role in quantum entangled cluster state generation, extreme environment sensing and dark matter detection.}
		\label{fig:future_directions}
	\end{figure}

	The advances in material processing and device integration are bringing new opportunities for fundamental research in color center platforms. We discuss implications to Bell inequality tests, multipartite entanglement, understanding of matter phase transitions at high pressures, geochemical explorations of deep-Earth conditions, and dark matter detection, as illustrated in Figure \ref{fig:future_directions}.
	
	\subsubsection{Further rejections of local-realistic descriptions of nature}
	While the diamond NV$^-$ center played a crucial role in the seminal demonstration of the loophole-free Bell test \cite{Hensen2015}, \deleted[id=SM]{the} \added[id=SM]{other} color center systems can be exploited to provide an even stronger inequality violation. The key to this approach is in increasing the number of tested events, and improving on the demonstrated \emph{one per hour} successful entangling rate.
	
	The first of the major sources of the indistinguishable photon loss between the two NV$^-$ centers originates from the low Debye-Waller (DW) factor, which leads to 95\% of light in each of the emitters being wasted to phonon interactions. The efficiency of this process can be improved either by integrating the NV$^-$ center into a photonic cavity or by using a different color center. Another major source of photon loss is the fiber absorption, which could be overcome by using a color center emitting at NIR wavelengths.
	
	The photonic integration route would provide up to two orders of magnitude more photons in the ZPL, resulting in up to four orders higher frequency of entangling events, bringing them to approximately \emph{one per second}. In this case, the null-hypothesis test could be reduced from $P=0.039$ to $P=0.001$ in order of hours, as opposed to months.
	
	The color center change would preferably require a system which exhibits ODMR contrast, has a high DW factor, and operates at fiber-friendly wavelengths. While not exactly at a telecom band, silicon vacancy (V$_{Si}^-$) and divacancy (V${Si}$V$_{C}^-$) in 4H-SiC both operate at a more desirable set of wavelengths ($\sim$900 nm and $\sim$1,100 nm) inducing two to four times higher transmission, and exhibit higher DW factors \added[id=MR]{\cite{udvarhelyi2020vibronic, koehl2011room}} than the diamond NV$^-$ center \added[id=MR]{(8-9\% and 7\% vs. 5\%)}\deleted[id=MR]{(10\% and 7\% vs. 5\%)}. The main research challenge with novel color centers is bringing their spin operation fidelities to levels comparable to the NV$^-$ center.
	
	Testing of Bell-type like inequalities is also foreseeable with color centers for potentially stronger violations of hidden variable theory predictions \cite{van2005statistical}. These include color centers with higher spin serving as qudits \cite{Nagy2019}, as well as the generation of Greenberger-Horne-Zeilinger (GHZ) states \cite{greenberger1989going} discussed in the next section.

	\subsubsection{Further explorations into quantum entanglement}
	
	\deleted[id=SM]{Quantum entanglement has}\added[id=SM]{Optical and microwave experimental schemes have} been utilized in color centers to entangle electronic spins and photons \cite{Hensen2015}, as well as electronic and nuclear spins \cite{nguyen2019integrated}. The next big frontier is the demonstration of photon-photon entanglement to construct cluster \deleted[id=SM]{entangles} \added[id=SM]{entangled} states \cite{lindner2009proposal}.
	
	As has been proposed \cite{economou2016spin}, the rich energy scheme of SiC silicon vacancy and nitrogen vacancy centers hosts the so called I-I transitions which contain two optical spin-preserving paths with orthogonal photon polarization that can be brought into a superposition. Subsequent optical pumping of such a system would provide an N-photon GHZ state, while the additional microwave or optical pulse interspersing sequence would create a quantum cluster state with arbitrary phase relations between qubits. Proof of principle experiments have been demonstrated with quantum dot biexcitons \cite{schwartz2016deterministic} entangling five photons, however, the coherence time in these systems is limited to $\sim$ 100 ns. Coherence time in SiC color centers is up to five orders of magnitude higher, providing opportunity to significantly grow the number of generated cluster entangled photons.
	
	Another route to cluster state entanglement is the application of time-bin entanglement to color centers integrated into a cavity \cite{lee2018towards}. Here, the ground states of a lambda system are brought into a superposition and optically pumped. The Purcell enhancement biases the optical transitions in the system to one arm of the lambda scheme. A subsequent $\pi$-pulse and an optical pumping cycle bring the emitted photon into a superposition of an early and the late time-bin, while the unknown collapsed spin state of the lambda system maintains the phase relation to the photon emitted in the repeated round of this process. This process has been utilized in quantum dots to generate the W-state or three entangled photons \cite{lee2018multi}. With the previously discussed advances in photonic device integration, this protocol can be implemented with color centers. 
	
	Minimizing decoherence of color centers is vital for the control of complex entangled states.  Advanced microwave pulse sequences have been developed using Hamiltonian engineering to cancel the effect of different types of many-body interactions \cite{ZhouLukin2019}.  For higher spin color centers (such as those in Figure \ref{fig:colorcenters}), pulses at multiple frequencies corresponding to higher order coherences may be utilized to further isolate color centers from noisy environments \cite{lei2017decoherence}.
	
	\subsubsection{Matter at extreme pressures}
	
	Both DC and AC field sensing using color centers offer unique opportunities to investigate matter at extreme pressures in diamond anvil cells (DACs).  DACs can reach quasi-hydrostatic pressures in the range of 5 - 500 GPa, but the sample space is limited to nL to pL. Conventional SQUID-based magnetometry requires careful background subtraction of the magnetization of the pressure cell, which can be orders of magnitude larger than such small samples, limiting the effectiveness of such a technique \cite{TaufourPRL}.  Conventional NMR is also precluded because the number of nuclei in these volumes is insufficient to detect \cite{Meier2018}.  Recently, NV$^-$ centers in diamond have been utilized to detect DC magnetic fields in DACs \cite{Hamlin2019}.  The diamond anvils are transparent to the excitation and fluorescent light, and the NV$^{-}$ centers can be located either on a chip within the sample space or directly on the culet of the anvil \cite{Lesik2019,Yip2019}.  Microwaves can be introduced via antennas directly located within the diamond culet \cite{NVinDAC}.  Because the zero-field splitting of the NV$^-$ center is sensitive to strain, it can be used both as a pressure manometer as well as a method to probe strain inhomogeneity within the DAC \cite{Hsieh2019}.  This technology offers an important avenue to investigate  phase transitions of correlated electron materials and \emph{magnetic} properties of materials at high pressures, such as room temperature superconductivity in hydrogen-rich materials \cite{Drozdov2019,LaHsuperconductivity} and metallization of hydrogen above 450 GPa \cite{Dias715}.   A challenge for NV$^-$ magnetometry is the collection of light through the anvil due to the high index of refraction of diamond.  Improved materials processing, such as the use of active photonic devices and microfabrication on the anvil may significantly improve the performance.  SiC anvils (Moissanite)  are often used instead of diamond due to their lower cost, especially at lower pressures where larger sample volumes are required, and multiple color centers are available in this case (see Table \ref{tab:1}). 
	
	\subsubsection{Future solution chemistry work}
	
	The exquisite sensitivity of NV$^-$ centers to AC magnetic fields enables the possibility of performing high resolution NMR spectroscopy at high pressures, which has the potential to fundamentally transform Geochemistry. 
	Surprising new reactions are being discovered via geochemical models at the high pressures and temperatures of the entire crust and upper mantle, such as precipitation of diamonds from water \cite{SverjenskyHuang2015}, but they extend to conditions beyond where direct experimental verification is possible. For example, the models use a computational estimate of the static permittivity of water as a central variable, because the high temperatures (up to $1200^{\circ}$ C) and pressures (up to 5 GPa) can't be approached experimentally\cite{Sverjensky2019}.  As discussed above, NV-based NMR spectroscopy can measure vanishingly small numbers of nuclei compared to conventional NMR. Therefore, direct testing of the geochemical predictions would be possible if spectra for solutes could be measured via defect-based optically-detected NMR spectrometer combined with diamond-anvil cells.
	
	Once the volume limitation of standard high-pressure NMR is eliminated, advances cascade rapidly.  For example, the phase diagram of water is highly distorted by some electrolytes at deep-Earth conditions \cite{JOURNAUX2017, JOURNAUX2013}, but the extents are virtually unknown --- it is clear that some salts can nearly double the freezing pressure of water \cite{Ochoa2015}.  The presence of liquid water controls the mineralogy created during plate subduction in the Earth and the generation of melts, which appear at the Earth surface as volcanoes.  Knowing the pressure limits of freezing would better constrain the conditions where water-based extraterrestrial life could be expected. Similarly, this technology will invigorate the study of dangerous or precious solutes.  Studying the solution chemistry of transuranic elements, for example, requires extraordinary safety precautions if the volumes are milliliters or greater, but the handling becomes simple if the volumes are nanoliters.  Sensing, particularly of magnetism, in remote locations becomes feasible if fluorescent nanodiamonds are used as in-line magnetometers in a microfluidics array, or if they are injected into a cell.  Magnetic resonance spectroscopy becomes possible in geometries that were previously unimaginable when detection was via induction coils.
	
	\subsubsection{Dark matter detection}
	
	In the late 1970's several galaxy rotation curves were measured and found to exhibit unusual rotational properties that could not be explained given the observed distribution of matter based on luminosity \cite{Rubin1980}.  These results provided the first evidence for dark matter, which has mass and interacts gravitationally, but only interacts weakly, if at all, with photons.  Dark matter is now an integral component of the standard cosmological models, and comprises more than 25\% of the known mass in the universe \cite{Planck2016}.  There have been numerous theories proposed as to the nature of dark matter, over a range of particle masses (rest energies), and searching for direct experimental evidence of dark matter is a major endeavour in high energy physics.  One class of theories that has attracted attention  posits  that dark matter consists of excitations of an axion field in quantum chromodynamics (QCD) \cite{AxionReview}.  The existence of these particles can potentially resolve certain mysteries, such as the appearance of  unexpected symmetries in the QCD Lagrangian (the so-called strong-CP problem).  Axions particles are expected to \deleted[id=NC]{an} experience a weak effective magnetic interaction with magnetic dipoles in matter \added[id=NC]{\cite{axionexptPRL}}, and several efforts have sought to identify their signature through magnetic resonance techniques.  The ADMX experiment utilizes a tuned microwave cavity and looks for excitations of the electromagnetic field due to interaction with axions  \cite{ADMX2001,ADMxPRL2018}. The CASPEr experiment aims to detect the precession of a nuclear spin in response to local axion dark matter field using a sensitive quantum interference device (SQUID) magnetometer to detect the time-dependent magnetization of the nuclei \cite{CASPEr2014,Garcon2017}.   The high magnetic field sensitivity of NV$^-$ centers in diamond might also be used to detect such fields \cite{Rong2018}.  An alternative use of color centers is to detect the strain induced in the crystal in response to the damage created by nuclear recoil due to scattering of a dark matter particle \cite{SushkovDarkMatter}. Importantly, the latter approach would enable directional detection, and is applicable to a broad range of potential dark matter candidates.  This approach takes advantage of the coupling between the zero-field splitting parameters of the color center and the local strain in the crystal.

	\section{Conclusion}
	Color center platforms have made significant strides in our understanding of fundamental science, from the tests of local realism, to the understanding of phase transitions in solid state systems. While most of these results have been based on diamond NV$^-$ centers, a host of emerging color centers and their integration with nanophotonic devices are poised to expand, as well as radically innovate, the toolbox for extreme environment sensing, generation of arbitrarily complex entangled states and dark matter detection. Beyond this, nano-engineering of color center platforms is also a path to the commercialization of these high-performing systems which would provide broad access to the cutting edge science discovery.

	\section*{acknowledgments}
	The work was supported by the United States Department of Energy, Office of Basic Energy Sciences, Chemical Sciences, Geosciences and Biosciences Division for Grant DE-FG0205ER15693.

	% \section*{conflict of interest}
	% You may be asked to provide a conflict of interest statement during the submission process. Please check the journal's author guidelines for details on what to include in this section. Please ensure you liaise with all co-authors to confirm agreement with the final statement.

\bibliography{Colorcenterreview_bib}

	%\begin{biography}[example-image-1x1]{A.~One}
	%Please check with the journal's author guidelines whether author biographies are required. They are usually only included for review-type articles, and typically require photos and brief biographies (up to 75 words) for each author.
	%\bigskip
	%\bigskip
	%\end{biography}
	
	%\graphicalabstract{future_directions2.eps}{This review showcases the fundamental discoveries across a variety of disciplines enabled by the color center platforms. The progress in material processing and the nanofabrication of photonic devices in diamond and silicon carbide are presented as tools for deepening the understanding of quantum physics, superconductivity and geochemical phenomena.}
	
\end{document}